# A Practical SAFE-AI Framework for Small and Medium-Sized Enterprises Developing Medical Artificial Intelligence Ethics Policies

Ion Nemteanu[1], Adir Mancebo Jr.[2], Leslie Joe[2], Ryan Lopez[2], Patricia Lopez[2], Warren Woodrich Pettine[3,4,*]

[1]Nemsee, LLC., Escondido, California
[2]The Data Science Alliance, San Diego, California
[3]Mountain Biometrics, Inc. (doing business as Medical Timeseries Networks), Salt Lake City, Utah
[4]Department of Psychiatry, The University of Utah, Salt Lake City, Utah
*Direct correspondence to warren.pettine@hsc.utah.edu

# Abstract

Artificial intelligence (AI) offers incredible possibilities for patient care, but raises significant ethical issues, such as the potential for bias. Powerful ethical frameworks exist to minimize these issues, but are often developed for academic or regulatory environments and tend to be comprehensive but overly prescriptive, making them difficult to operationalize within fast-paced, resource-constrained environments. We introduce the Scalable Agile Framework for Execution in AI (SAFE-AI) designed to balance ethical rigor with business priorities by embedding ethical oversight into standard Agile-based product development workflows. The framework emphasizes the early establishment of testable acceptance criteria, fairness metrics, and transparency metrics to manage model uncertainty, while also promoting continuous monitoring and re-evaluation of these metrics across the AI lifecycle. A core component of this framework are responsibility metrics using scenario-based probability analogy mapping designed to enhance transparency and stakeholder trust. This ensures that retraining or tuning activities are subject to lightweight but meaningful ethical review. By focusing on the minimum necessary requirements for responsible development, our framework offers a scalable, business-aligned approach to ethical AI suitable for organizations without dedicated ethics teams.

# Introduction

AI systems deployed in healthcare, whether directly in clinical decision-making or through supporting technologies, carry significant ethical responsibilities. Their outputs can influence diagnoses, treatment options, operational workflows, and ultimately patient well-being. For organizations developing or integrating these tools, including small and medium-sized enterprises (SMEs) that support patient care, ethical considerations such as fairness, transparency, and accountability are essential from the outset.

While significant academic work has explored the ethics of deploying artificial intelligence (AI) in medical decision support systems, these insights are often inaccessible to small and medium-sized enterprises (SMEs) (Ali et al., 2023; Crockett et al., 2023; Hajianhosseinabadi & Lindau, 2024; Kostick-Quenet et al., 2022; Ning et al., 2024; Winecoff & Watkins, 2022). The complexity of AI ethics, including concepts like fairness, transparency, and accountability, presents a significant barrier, as does the prohibitive cost of dedicated ethics teams. Furthermore, many existing toolkits lack the practical guidance or real-world case studies necessary for SMEs (Fjeld et al., 2020). Additionally, embedding ethics into AI development can be transformative for SMEs. Investments in ethical governance contribute to the value of intangible assets such as intellectual property, loss aversion, brand reputation, and organizational goodwill. In healthcare markets, where regulatory trust and patient safety are paramount, demonstrable ethical alignment can also serve as a differentiator - attracting ethically aware customers, clinical partners, and investors. As a result, operationalizing AI ethics is not merely a compliance strategy but a value-creating asset for emerging technology companies. Proactive approaches tend to emphasize ways in which investments drive important intangibles that are reflected in a differentiated reputation and robust corporate culture. Mature firms link these intangible outcomes directly to investments in AI ethics and governance, using criteria such as recognitions of industry leadership, support of Environmental, Social and Governance (ESG) efforts, as well as the long-term ability to manage risks (Domin et al., 2024).

In April 2024, the Department of Health and Human Services (HHS) Office for Civil Rights (OCR) issued a rule under Section 1557 of the Affordable Care Act that explicitly prohibits discriminatory outcomes from patient-care decision-support tools, including AI. The rule specifically references the use of clinical algorithms in decision-making, clarifying that these tools range from flowcharts and clinical guidelines to complex computer algorithms, decision support interventions, and models (Centers for Medicare & Medicaid Services, 2024). However, no explicit federal regulations directly govern the broader development and deployment of AI in healthcare, particularly outside the narrow scope of Software as a Medical Device (SaMD). This leads to significant ambiguity in compliance requirements and risk management, especially as ethical, legal, and operational challenges emerge. While the FDA oversees certain AI tools through its existing medical device pathways, and the HHS Office for Civil Rights has issued limited guidance on nondiscrimination in automated systems, the broader regulatory environment remains unsettled. Notably, recent legislative proposals such as the "One Big Beautiful Bill Act" seek to impose a 10-year moratorium on state-level regulation of AI systems engaged in interstate commerce - essentially preempting most sub-federal efforts to introduce stricter oversight. While exceptions exist for laws that *facilitate* AI adoption or enforce general-purpose provisions, the bill underscores how contested and politically sensitive AI regulation remains in the U.S. context. The result is a fragmented and uncertain landscape in which developers must anticipate future federal frameworks while contending with limited, and sometimes conflicting, present-day oversight. Thus, the prevailing consensus in policy circles remains that the question is not *if* stricter regulation will arrive, but *when* - and what form it will take.

To facilitate SMEs in addressing this challenge, we introduce a simplified framework for responsibly integrating artificial intelligence (AI) in medical settings. Utilizing this framework will help ensure SMEs develop medical AI solutions that meet the standard of large hospital systems and prioritize patient care (Coalition for Health AI, 2023).

Emphasizing AI responsibility, our framework addresses fairness, transparency, and ethical deployment for all individuals. A well-defined process facilitates collaboration between healthcare practitioners and data scientists, ensuring that ethical considerations, fairness, and equity are systematically assessed and documented. This approach ensures that each phase of AI development aligns with these critical values, leading to more effective and responsible AI tools for healthcare.

# Methods

### Study Design and Project Context

This work followed a design-science, practice-oriented methodology to develop **SAFE-AI**, a scalable internal governance framework intended for broad adoption by small-to-medium enterprises (SMEs) building medical-AI products. The project was a collaboration among Medical Timeseries Networks (MTN; an SME), the Data Science Alliance (DSA), Nemsee LLC, and the University of Utah. MTN served as the primary use-case organization, while DSA, Nemsee, and academic partners contributed ethics, regulatory, and methodological expertise.

### Informal Evidence Gathering

Between December 2024 and April 2025, the design team conducted an ad-hoc search of peer-reviewed literature, regulatory guidance (eg, FDA SaMD Good Machine-Learning Practice, HHS §1557 rule, EU AI Act drafts), and industry white papers. Searches were performed opportunistically in PubMed, Scopus, IEEE Xplore, and grey-literature repositories as questions arose during development. All relevant sources were captured in a shared Zotero library and tagged by topic. Key findings were reviewed in weekly design-committee meetings.

### Iterative Co-design Process

Framework creation unfolded over 20 weeks (January–May 2025) through an Agile, design-thinking cycle:

1. **Initial workshop (2 h):** a structured needs-finding session with MTN's data-science team established pain points and success criteria.
2. **Weekly design meetings (1 h each):** one MTN executive, three DSA AI-ethics researchers, one Nemsee consultant, and one University of Utah professor met to prototype, critique, and refine SAFE-AI artifacts.
3. **Embedded expertise:** DSA working group itself functioned as a collective body of domain and data science expertise, drawing on extensive project experience across healthcare, industry, and public-sector AI deployments. This structure enabled ongoing input from practitioners without the need for separate interview protocols, allowing design insights to emerge continuously through direct participation.
4. **Documentation:** meetings were recorded on Zoom; transcripts were summarised by an AI agent.
5. **Iteration:** the team completed three full prototype→feedback→revision cycles, each lasting ~6 weeks.

### Framework Outputs

The co-design process yielded:

- a four-phase life-cycle diagram (**Discovery, Assessment, Development, Monitoring**);
- stage-specific checklists covering acceptance, fairness, and transparency metrics;
- practical implementation guidance (eg, role assignments, artefact templates) for each phase.

### Consensus and Documentation

All design decisions were made through real-time discussion until unanimous verbal agreement was reached; no formal voting or Delphi procedures were used. Version history, meeting minutes, and rationale for changes were maintained in a locked Google Drive folder that serves as the project's audit trail.

### Planned Pilot Implementation and Evaluation

SAFE-AI will be applied to two upcoming MTN initiatives - a patient-monitoring product and a data-engineering pipeline. A lightweight evaluation plan was pre-specified to capture:

- **Process metrics:** time (hours) required to complete each SAFE-AI phase.

- **Governance metrics:** count and category of fairness metrics defined per project.
- **Usability:** post-implementation survey of participating staff (5-item Likert scale; 1 = strongly disagree, 5 = strongly agree).

Descriptive statistics will summarise these data after the first pilot quarter.

### Ethical Considerations

The project involved policy development only; no human participants or identifiable health data were used. Institutional Review Board (IRB) oversight was therefore not required.

# SAFE-AI Ethical Evaluation Framework

We present the Scalable Agile Framework for Execution in AI (SAFE-AI) - a structured process designed to embed ethical evaluation into the technical development lifecycle of AI products (summary in Fig. 1, full process in Fig. 2). SAFE-AI enters the workflow after user stories, business requirements, or system objectives have been defined and prioritized, and tasks are ready for execution by the development team. It complements Agile and Scrum methodologies by introducing structured ethical oversight at critical stages without disrupting project momentum.

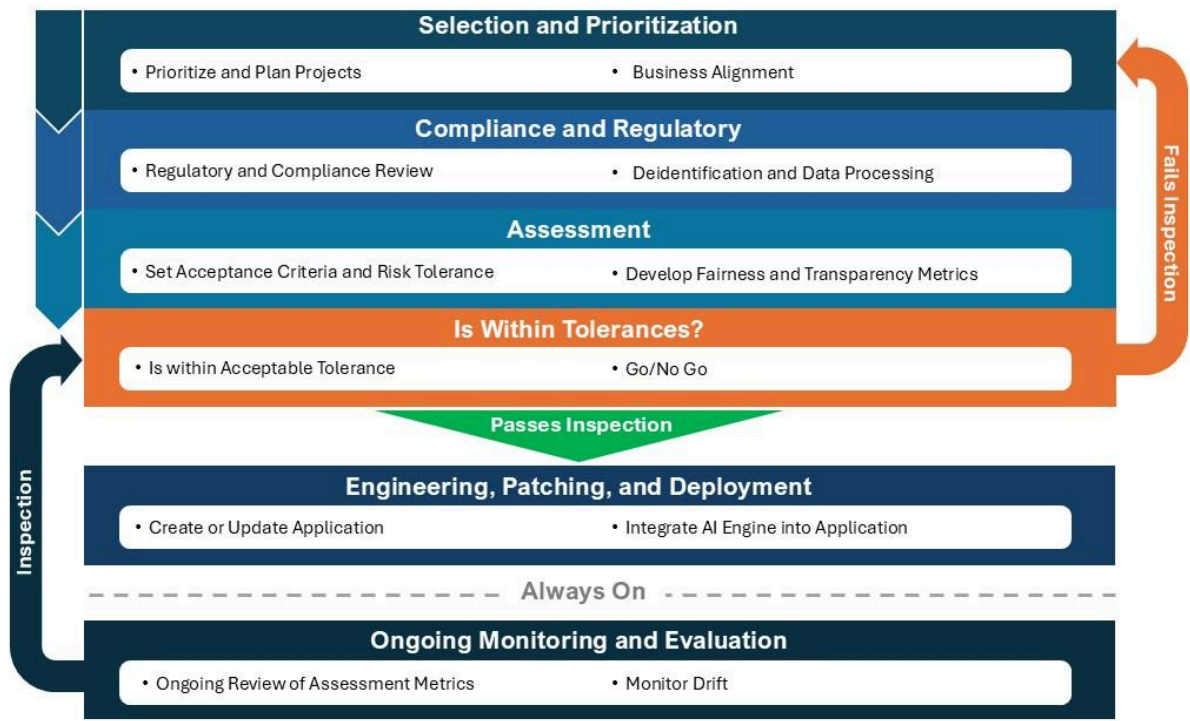

**Figure 1**. SAFE-AI Summary Workflow - highlights each core phase and feedback loop, from prioritization to deployment and monitoring.

The SAFE-AI framework is not a one-time gate. It is designed to be re-entered at any point in the development lifecycle whenever changes occur that could materially affect the behavior, fairness, or transparency of the AI system. This includes retraining with new data, hyperparameter adjustments, updates to feature selection, or integrating external datasets.

Practitioners are encouraged to revisit SAFE-AI whenever there are any modifications to their models, regardless of perceived simplicity. Even seemingly minor technical tweaks may cause unintended shifts in model performance or equity. Further recommendations are outlined in the section titled "Quick Note on Model Updates and Considerations," where we suggest re-evaluating all previously established metrics (see Part II – Assessment) to ensure continued ethical integrity.

By embedding lightweight, repeatable ethical checkpoints into natural inflection points, SAFE-AI enables rapid iteration without sacrificing responsible oversight. It provides a scalable governance framework that is especially suited to small and medium-sized enterprises looking to operationalize ethical AI in a practical, execution-ready manner.

In this paper, references to a *"data science team"* is used as a generalized term to encompass data scientists and the other skilled roles and internal stakeholders that contribute to the project including product and program managers, data engineers, analysts, statisticians, software engineers, and DevOps personnel. However, when referring specifically to data scientists, we mean those roles that are directly responsible for researching and building the AI algorithms for use in the business and in their products.

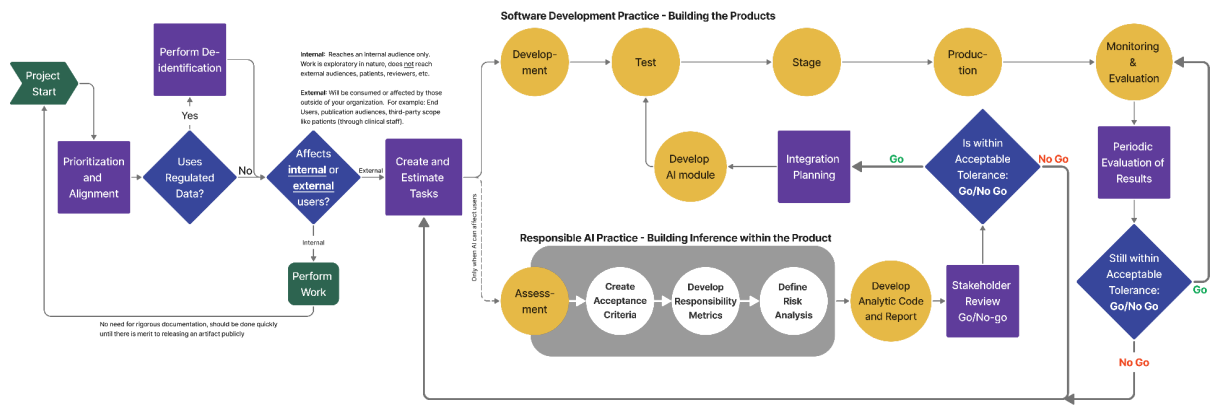

**Figure 2.** Detailed Ethical Evaluation Process SAFE-AI - conducted continuously throughout product development as seen above. The process begins with a discussion of priorities and alignment and proceeds through model building and inference implementation, emphasizing appropriate tolerance levels at each iteration.

# SAFE-AI: Part I - Discovery

**Determine if the Project Will Use Regulated Data?**

First determine if the planned work will use or access regulated data. Organizations must comply with all relevant regional, state, federal, and local laws <u>before</u> any access and manipulation of data begins. Protected Health Information (PHI) and Personally Identifiable Information (PII) are among the many regulated data types. Existing acts such as the General Data Protection Regulation (GDPR) and the California Consumer Privacy Act (CCPA) set specific compliance requirements.

Best practices include anonymizing or aggregating data to protect individual privacy. Additionally, informed consent is required for legally effective participation in research and must be obtained in accordance with the principle of respect for persons (Informed Consent FAQs, n.d.). This principle ensures that individuals are given the opportunity to choose whether they wish to participate in research studies.

A comprehensive regulatory review should be conducted before any AI development begins. If specific handling techniques require documentation, they should be securely stored and referenced throughout development. When designing systems that impact external audiences, regulatory requirements should be listed as acceptance criteria and revisited at every development stage. If the same data are used repeatedly, the regulatory review can cover repeated use and should have a documented process (if necessary) to ensure compliance for future access and proper use. This will ensure that resources do not have to review the data every time it is accessed.

**Determine if Scope of Work Impacts Internal or External Users?**

During typical prioritization meetings, it should be discussed if the work will affect external stakeholders like patients, clinicians, or other healthcare providers. This is an important distinction since there are many tasks that should be done quickly and match the pace of the organization and do not require the extra burden of documentation. Realistically, not every task requires burdensome processes and documentation. This framework aims to only target the tasks that have a direct effect on the healthcare outcomes as a direct result of the product being developed. The first step in this process is a quick discussion during the planning meeting that determines the upcoming scope of work for the development team and to clarify if outputs can affect external stakeholders. Clarifications are as follows:

> **Internal Stakeholders:** Exploratory research, feasibility studies, or internal management tools that do not directly impact patient outcomes can proceed without additional scrutiny.

> PROCESS: In this case, move forward with the task as normal. No need for additional scrutiny. Once complete, start the process over from the beginning.

> **External Stakeholders**: Work that affects those outside of your organization. For example: patients, clinicians, or other healthcare providers; end users, publication audiences. In this case, the outcome of these tasks can or will have a direct impact on an external audience.

> PROCESS: In this case, move on to the next step in this process.

**Create and Estimate Tasks:**

This phase begins with breaking down high-level project goals into smaller, manageable user stories or tasks. These tasks often span data ingestion, cleaning, feature engineering, model development, evaluation, and integration. During planning, the data science team estimates each task's duration based on the scope of work for iterative delivery while leaving room for learning, model iteration, and integration into production workflows.

To ensure that ethical and social considerations are not deferred until the end of development - or left out entirely - this framework incorporates structured ethical reasoning into the task creation process from the start. The foundation for this integration draws from the **RESOLVEDD** process, a structured ethical decision-making framework originally introduced by Pfeiffer & Forsberg (2014). RESOLVEDD guides practitioners through steps such as Reviewing the facts, Estimating the ethical conflict, Surveying alternatives, Opening options to stakeholders, Listing risks, Valuing impacts, Evaluating options, Deciding, and Defending the decision. While originally applied to workplace ethics, the logic of RESOLVEDD has broad applicability in responsible AI contexts.

Building on this foundation, **Vakkuri et al. (2021)** introduced **ECCOLA**, a card-based toolkit tailored to agile AI and autonomous systems development. ECCOLA does not follow an acronym, but its sprint-by-sprint format maps well to the RESOLVEDD process. It consists of a deck of 21 cards aligned to 8 core AI ethics themes (e.g., transparency, accountability, bias), each designed to provoke discussion, surface risks, and frame decisions during active development. The cards can be used in planning meetings to ensure that ethical issues are not only acknowledged but actively scoped into the project as discrete tasks.

By aligning ECCOLA cards to user stories, teams can define **"responsibility metrics"** alongside traditional performance metrics. For instance, if a card on transparency is selected, the team may create a task to design interpretability strategies for end users. If a card on bias is relevant, a corresponding task might include fairness audits or subgroup performance analysis. In this way, ECCOLA serves as a lightweight but structured extension of the RESOLVEDD reasoning model, embedding ethical reflection directly into the software development lifecycle (SDLC) in a format that scales for both large enterprises and lean startups.

This approach helps convert abstract ethical principles into concrete, time-boxed actions that can be tracked and revisited throughout the sprint cycle. It ensures that responsibility is not just a post hoc evaluation, but a proactive part of scoping and estimating work. The following sections explore how this integration supports transparency, interpretability, and stakeholder trust through novel metrics and communication methods.

## SAFE-AI: Part II - Assessment

> **Box 1: Defining distinct task categories that are relevant to the SAFE-AI process**
>
> **Developing Product Engine (PE):** Traditional software development where the software product and architecture are built. The PE involves building systems that are highly repeatable, highly predictable, and easily tested. For example, building a database is predictable because there is no mathematical uncertainty involved - tables are either created or not, fields either exist or they don't, and data either matches a schema or it doesn't. It is clear when the task is complete, and the outcome is deterministic. A database is a database, features are features - done. Similarly, building a system that calls data from that database and presents it in an application is also straightforward. A software team can build the product and create a holding spot for the data that will be produced by the inference engine. A successful product typically runs the same code twice and gets the exact same result.
>
> **Developing an Inference Engine (IE):** On the surface, an inference engine may look similar to traditional product features. It has code, it executes functions, and it reads from and writes to databases. However, where it differs is significant. Unlike product code - which is deterministic and repeatable - the inference engine introduces mathematical uncertainty. The data it processes and outputs often relies on probabilistic models, randomization, or sampling techniques. This means running the same code twice can produce different results, and that's not a bug - it's an expected behavior. While a database or UI element has a clear "done" state (e.g., a schema exists or it doesn't), an inference engine outputs predictions that must be interpreted within a range of acceptable outcomes. There's no binary pass/fail: instead, performance must be evaluated across multiple runs, data splits, and use cases.
>
> This makes the IE phase inherently more complex to test. It requires rigorous evaluation from data scientists to ensure model predictions fall within the acceptable predefined thresholds, and that edge cases or subgroup performance are well understood. Validation includes both technical accuracy and conceptual soundness and proving that the model(s) behave as intended under typical and exceptional conditions.

As a result, organizations often manage two distinct codebases with different versioning cadences. One includes the core functional product code that will be exposed to end users. The other is research-oriented code maintained by the data science team—consisting of modeling notebooks, experimental logic, hypotheses, tuning cycles, and visualizations. This research code must undergo thorough validation before being promoted. Only when the team agrees that the model meets uncertainty and risk criteria does it become eligible for integration.

**AI Integration (AII):** TThis is the convergence point. AI Integration (AII) is where a validated inference engine is embedded into the product codebase. This integration code is functional in nature: it accepts input, runs the algorithm, and outputs predictions. It's optimized for performance, often stripped of exploratory elements, and ready for deployment. While implementation methods vary, the AII phase follows a consistent three-stage flow: external systems supply input, processing and prediction occur, and output is routed to the product interface.

By the time the code reaches this phase, the model has been fully validated and is no longer being evaluated for feasibility—just functionality. The code is versioned independently and integrated in coordination with product releases to ensure seamless deployment.

To illustrate these differences, Figure 3 below demonstrates how these phases operate in practice:

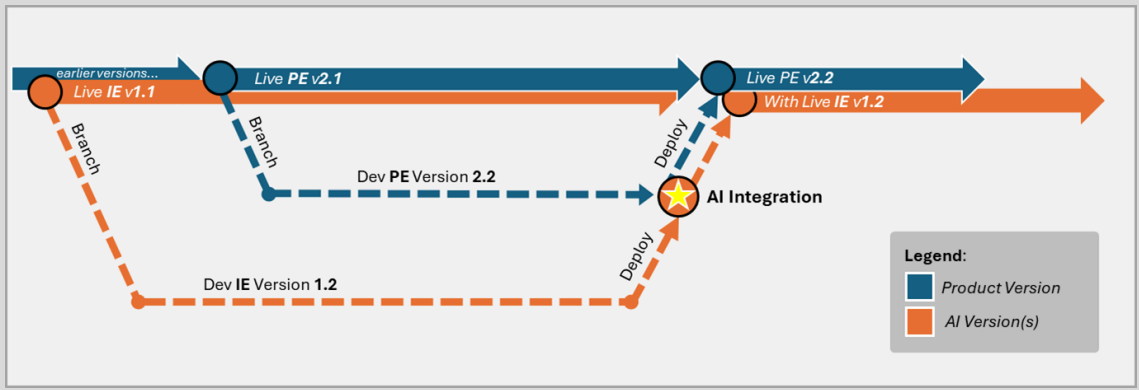

**Figure 3.** Example of PE and IE version control with parallel development cycles.

As shown, the Inference Engine (IE) and Product Engine (PE) evolve on separate but parallel development tracks, each with its own branching and deployment cadence. For example, Live IE version 1.1 may remain in production while a newer Dev IE v1.2 is being validated, just as Dev PE v2.2 advances independently of earlier releases. The AI Integration (AII) phase serves as the coordination point where validated AI models are embedded into product-ready systems and inputs and outputs are aligned and locked.

> This moment—marked by the star icon—represents the culmination of both tracks, where a stable inference engine is paired with its corresponding product version for release. Once deployed, the system enters a new live state (e.g., Live PE v2.2 with Live IE v1.2), ensuring that only approved, tested combinations reach end users. This disciplined integration workflow supports transparency, traceability, and accountability across the AI development lifecycle.
>
> It's important to note that changes to the Inference Engine (IE) don't always involve modifications to input or output formats. The structure of the data pipeline may remain consistent even as the underlying model evolves—through algorithm updates, parameter tuning, or other enhancements. The purpose of the diagram is simply to illustrate that, at some point, the IE and Product Engine (PE) must converge during the AI Integration phase. However, since IE updates can occur independently of PE deployments, it's critical that any such changes—regardless of how minimal they may appear—still undergo the SAFE-AI evaluation process. This ensures that even behind-the-scenes model improvements are held to the same standards of validation, risk assessment, and accountability.

This section occurs during the Inference Engine (IE) phase and outlines key steps the data science team - both technical and product stakeholders - should complete before progressing to development and integration. These steps ensure that AI inference engines operate within defined boundaries of uncertainty and acceptable risk, aligning with ethical standards and strategic business goals.

It is important that the following steps be incorporated alongside traditional software testing as a routine part of development practice, with consistent attention to their execution. While these steps may initially appear to add time or complexity - requiring deeper evaluation, more test cycles, and added stakeholder input - the SAFE-AI framework is designed to be iterative and collaborative, ensuring that ethical assessments are embedded incrementally and transparently throughout the development lifecycle. Crucially, tasks are not imposed in isolation; they are co-developed and agreed upon by all stakeholders, including data scientists, engineers, product leads, and domain experts, which keeps timelines aligned and priorities clear.

When planned appropriately, these activities do not disrupt schedules - they become part of the schedule. Ethical testing is treated with the same rigor and visibility as functional testing, making it easier to forecast and manage. While often overlooked due to perceived resource constraints or competitive pressure, these steps are essential to building systems that are robust, compliant, and trustworthy. The framework's goal is to define the minimum necessary ethical safeguards required to ensure responsible AI, while preserving agility and long-term strategic value.

There are three tasks that are especially relevant during the assessment phase, each of which are detailed below:

**Creating Acceptance Criteria:**

Acceptance criteria define the conditions that a product or feature must satisfy before they are considered complete. The concept is borrowed from established practices in traditional software development lifecycles (SDLC). For example, in the development of AI-powered clinical diagnostic tools, acceptance criteria may include achieving a minimum accuracy threshold of 95% in identifying specific conditions, or demonstrating consistent performance across diverse patient demographics. Such criteria help establish shared expectations and support regulatory alignment before AI models are integrated into clinical workflows. Atlassian describes acceptance criteria as "clear, concise, and testable statements that focus on providing positive customer results" (Atlassian, n.d.).

In AI systems, especially those used in high-stakes domains like healthcare, defining acceptance criteria is not just a technical step - it is a normative one. The choice of what counts as "acceptable" performance can have real-world implications for safety, equity, and trust. It is therefore essential that these criteria be co-developed with key stakeholders, including clinicians, patients, domain experts, and end users. AI engineers may assume they understand what matters most to users, but this assumption often overlooks critical perspectives - particularly around risk tolerance, fairness, and real-world utility. Engaging stakeholders in defining success conditions helps ensure that the system reflects the values and priorities of those it is intended to serve.

While traditional software engineering often emphasizes binary outcomes ("pass/fail"), fairness in AI requires additional safeguards beyond standard SDLC practices. For instance, an algorithm can return a reliable "yes" or "no" and pass functional tests, but there may be no checks for how accurate those results are across demographic groups, or how transparently those outcomes can be interpreted. Embedding fairness, equity, and performance expectations into the acceptance criteria phase is one of the most direct and impactful ways to operationalize ethical principles in AI development.

**Develop Responsibility Metrics:**

Now that the acceptance criteria have been established, it is time to consider the range of impacts that algorithmic results may have on real individuals—including those who do not interface with the system directly but are still affected by its outcomes. For example, a doctor may use the algorithm's output directly to inform a clinical decision, while a patient may be affected by that decision indirectly. This section will discuss two distinct types of metrics that are both required in order to ensure responsible use of data: Fairness Metrics and Transparency Metrics.

**Fairness Metrics:** Define and measure bias within AI systems. As The Software Alliance (2021) states, fairness is subjective and requires selecting the most appropriate metrics for the system being developed. These metrics should be documented and reviewed collaboratively by subject matter experts and data scientists (The Software Alliance, 2021). In practice, fairness metrics are used to detect and mitigate disparate impacts—ensuring that one group or class (e.g., based on race, gender, age, or geography) is not disproportionately harmed or excluded compared to others. This may involve evaluating performance across subgroups, measuring disparities in false positives or negatives, or assessing equal access to beneficial outcomes. The goal is not only to identify statistical imbalance, but to actively choose metrics that reflect the ethical priorities and risk tolerances of the system's intended users and affected populations.

**Transparency Metrics:** Define metrics that focus on ensuring that AI model predictions are interpretable, explainable, and actionable by end users. One technique used in this framework is a scenario-based probability analogy mapping method - a structured approach to communicating AI predictions by framing them as outcome narratives grounded in real-world context. This method, which we refer to as scenario-based probability analogy mapping method (SPAMM), enhances transparency by mapping probabilities to familiar analogies that illustrate the likelihood of different scenarios, the conditions under which errors occur, and their potential consequences. Rather than focusing solely on model internals, SPAMM presents predictions in narrative form - conveying not only what the model predicts, but why, how alternative outcomes compare, and which inputs most influence the result. This approach builds stakeholder trust by making uncertainty and risk both visible and relatable**.**

This approach is particularly important in high-stakes domains like healthcare, where decision-makers such as clinicians, patients, or administrators must weigh the risks of false positives and false negatives. By highlighting the probability of misclassification and mapping those scenarios to familiar contexts, stakeholders can assess how often and under what circumstances the model may be wrong - and what the real-world consequences of those errors might be. For example, if a model is correct 85% of the time but consistently fails on a particular patient subgroup, this insight can guide mitigation strategies, prompt workflow adjustments, or justify additional human oversight. These dimensions of model behavior - error distribution, uncertainty, and impact - are often absent in standard AI evaluation pipelines.

Traditional interpretability tools like SHAP (SHapley Additive exPlanations), LIME (Local Interpretable Model-Agnostic Explanations), Partial Dependence Plots (PDP), and Accumulated Local Effects (ALE) are valuable for understanding model behavior. SHAP, based on cooperative game theory, assigns each feature an importance value representing its contribution to a particular prediction - similar to calculating how much each player contributed to a team's win. LIME works by perturbing the input data near a given instance and fitting a simple model to explain the local decision boundary. PDP and ALE methods help assess how the average model output changes when a single feature is varied, which supports global interpretability, especially when exploring relationships between inputs and outcomes.

These methods help data scientists and auditors understand what the model is doing, but they may not fully address the kinds of interpretive needs that arise when communicating with clinical or business stakeholders. Technical performance metrics that evaluate true and value positive rates such as RMSE, Precision, and/or Recall can be difficult to translate to non-technical audiences. As Ribeiro et al. (2016a) noted, trust in machine learning models depends on the user's ability to anticipate and understand model behavior in practical terms - not just accuracy, but reliability and risk.

As a result, we apply SPAMM, which adds a critical layer to the transparency toolkit by reframing predictive outputs in terms that mirror how people naturally reason under uncertainty. At its core, the method walks users through a model's output by anchoring it to the specific input conditions that led to that prediction—and by showing how often those conditions have historically produced similar outcomes. For example, if a model predicts outcome "C" among options A, B, and C, it may also convey that in 70% of similar cases, variables $x_1$, $x_2$, and $x_5$ were associated with outcome C, while in 20% of cases, the model predicted outcome B under slightly different input patterns. Presenting outputs in this context—grounded in familiar patterns and frequencies—helps stakeholders understand not just what the model predicted, but why it arrived at that decision, how likely alternative outcomes were, and how much confidence they can place in it. While this can become complex in models with high-dimensional input spaces or large numbers of interacting variables, the point is not to replicate the entire model logic—it is to provide a real-world testing lens that frames predictions through the lens of practical "what-if" scenarios. This helps surface critical edge cases, increase interpretability, and build trust—especially in settings where human oversight and risk-based decisions are essential.

Crucially, this framing also enables teams to assign tangible costs to errors by explicitly asking: What can happen if this prediction is wrong? and What is the risk impact to the organization or patient? For example, a false negative in a sepsis prediction model may delay critical treatment, leading to clinical harm, regulatory exposure, and reputational damage. A false positive in an insurance fraud detector may result in a wrongful denial of benefits, eroding user trust and increasing legal liability. By mapping these risks to predicted outcomes and associated variables, organizations can better understand which errors matter most, estimate their financial or operational consequences, and prioritize safeguards accordingly. Integrating these cost-of-error considerations into the transparency process transforms interpretability from a purely technical exercise into a tool for risk mitigation and strategic alignment.

Ultimately, this approach supports a more robust, stakeholder-centered view of transparency. It enables users to understand reliability, anticipate error, and make informed decisions based on modeled outcomes - an essential capability for ethical and accountable AI deployment in healthcare and other sensitive domains.

This framework invites further development in areas such as visualization, user-centered metrics, and storytelling techniques for AI transparency. While this paper does not formalize the full methodology, it highlights the importance of designing explanation strategies that reflect how real-world users evaluate uncertainty, reliability, and relevance - not just technical performance.

Any of these measurable approaches should be established through a collaboration between subject matter experts who know the population of users that could be affected by the product, and by the data science team that has an intimate understanding of the data that is available during the training of the model as well has how the model will learn against those data and how to measure the error effectively.

As described earlier in the task estimation phase, tools like ECCOLA can support the creation of project-specific responsibility metrics integrated into an agile software development framework such as the below card example found in Illustration 1:

| Analysis | Minimize Gender Bias |
|---|---|
| **Problem**: The algorithm could unintentionally predict better for one gender than another. Gender is a protected and regulated field for decision support systems in healthcare. ||
| **What to do:** | **Output:** |
| 1. Check the balance of the raw training data generally and across all relevant features that could affect the training of the model and the models predictions. | 1. Distribution analysis of records per gender<br>2. Distribution analysis of gender records per features |
| 2. Check the balance of gender during the randomization of the training sets that are used on the fly to train the model. | 1. Distribution analysis of gender records (same as above) for each training set and proposed test & validation sets |
| 3. Check the predictive accuracy (based on declared error metrics) | 1. Distribution analysis of gender records (same as above) for each |
| after training for each gender and ensure they are balanced | training set and proposed test & validation set |
| 4. Check the distribution of impact based on final acceptance criteria test metrics | 1. Distribution analysis of gender records (same as above) for each final outcome as seen in the final product. For example: How many Males were "redirected to another clinic vs females" |

Illustration 1

**Define Risk Analysis:**

Risk analysis should be conducted collaboratively including consultation with relevant stakeholders like product teams, data scientists, and legal experts. There are many possible approaches and the following are a few examples of structured risk analysis processes that already exist in the field of AI and healthcare.
1) The NIST AI Risk Management Framework provides structured guidelines for evaluating risks in AI systems, emphasizing trustworthiness and resilience through iterative assessments (NIST AIRC, n.d.).
2) McKinsey's 'Derisking AI by Design' approach highlights the necessity of integrating risk management throughout the AI lifecycle, focusing on proactive monitoring and governance (Baquero et al., 2020).
3) Stanford's AI healthcare risk assessment framework outlines liability risks in medical AI applications, suggesting adaptive monitoring strategies to address unforeseen ethical and operational challenges (Mello & Guha, 2024).
4) The FDA's Total Product Life Cycle (TPLC) framework ensures continuous oversight from AI model development through post-market surveillance. The FDA writes how GenAI presents potential risks that may require varying levels of risk controls for different applications, as is true of other technologies. For example, hallucinations, particularly those that may appear to be authentic outputs to users, may present a significant challenge in certain health care applications where highly accurate, truthful information is critical. They state how current evaluation approaches, such as those used to evaluate computer-assisted triage, detection, and diagnostic devices, may still be applicable for GenAI-enabled devices, albeit with additional supporting evidence. However, it may be challenging to determine the evidence that may be needed for certain GenAI implementations (Health, 2023).

All of the above approaches include hazard identification, risk assessment, and mitigation strategies tailored to AI's dynamic nature in healthcare. By incorporating structured risk evaluation, teams can proactively address potential harms and regulatory challenges. Key considerations include:

- **Misuse Potential**: Could the product be used incorrectly, leading to harm?
- **Error Costs**: What are the consequences of false predictions?
- **Societal Implications:** Are there broader ethical concerns?

Legal experts should be engaged to ensure compliance with evolving regulatory landscapes and provide specific tests and metrics that can be used to create risk boundaries for AI models that can help pass regulatory reviews as well as build trust with the public.

### Quick Note on Model Updates and Considerations

Once an inference engine has been approved and integrated into a production environment, it is common practice to revisit the model for improvements through techniques such as retraining with new data, hyperparameter tuning, or feature modifications. While these changes may be perceived as incremental or routine, they carry the potential to significantly alter the algorithm's behavior, calibration, and fairness properties. Even minor updates can reintroduce latent biases, shift subgroup performance, or affect the interpretability and reliability of outputs.

From an ethical standpoint, any modification to the inference engine including changes to a single model or updates to the system of models should be treated as a new model iteration when following the SAFE_AI framework, regardless of whether the foundational architecture remains unchanged. Although it is not necessary to recreate the entire ethical evaluation pipeline, it is considered best practice to reapply the original assessment metrics established during initial development (outlined in Part II – Assessment). This allows teams to validate that the updated model continues to meet ethical expectations for fairness, transparency, and performance consistency.

In high-stakes domains such as healthcare, the tension between deployment speed and responsible oversight is a persistent challenge. Rapid iteration is often essential for competitiveness; however, this urgency must be weighed against the risk of unintended harm. Ethical oversight should not be perceived as a barrier, but rather as a safeguard that ensures continuity of ethical alignment across the model's lifecycle.

Therefore, organizations are strongly encouraged to adopt a policy of proactive reassessment following any material change to an AI system. By incorporating these evaluations into their standard operating procedures, teams can preserve ethical integrity, mitigate risks, and maintain regulatory and stakeholder confidence.

# SAFE-AI: Part III - Development and Integration

After defining the metrics and performing all the necessary reviews, the next step is to build and assess the model. This phase encompasses three key stages: **algorithm development, product development, and integration**. Each stage plays a crucial role in ensuring the AI system is both functional and effective within the healthcare setting.

**Algorithm development:** This initial phase focuses on creating a robust and efficient AI model tailored to a specific healthcare problem. It is heavily research-driven, involving hypothesis testing, feature engineering, model training, and validation. The primary objective is to establish a reliable algorithm that produces consistent and meaningful predictions. While coding in this phase is extensive, the outputs at this stage are primarily analytical and, in most cases, unusable. Throughout this phase, there will be charts and reports, model performance evaluations and metrics, and many times, code will be tested and fail with lessons learned as part of the scientific method. This research phase is analogous to developing a new mathematical formula, where the scientist is focused on testing its effectiveness and fine-tuning its parameters before considering real-world deployment. The goal of this process is to iterate on the system of models until an acceptable level of criteria has been achieved that balances accuracy against all the risks and value that can be attained by delivering the model. The resulting code is exploratory in nature and is not intended for direct integration into the final product.

**Product development:** Once the algorithm has been validated, the next step is embedding it within a functional, user-facing application. This phase includes two parallel development tracks:

1. **Application Development** – Designing a scalable, user-friendly software interface where the algorithm will operate. This phase works on setting up functional algorithms into the architecture to import the data, execute the algorithm, and output results in a way that is consumed by the product architecture. There are clearly defined inputs and outputs that are agreed upon between the algorithm developers and the product developers. This phase bridges theoretical research with practical engineering to create a scalable product that meets user needs.
2. **Algorithm Integration** – Translating the AI model into an executable function that can be triggered within the software environment.

**Integration:** Integration, both initial and ongoing, ensures that the AI model seamlessly interacts with the product's ecosystem. The integration phase ensures that the AI solution is not just theoretically sound but also operationally viable, meeting both technical and regulatory requirements before full deployment. This involves:

- Embedding the model within the product infrastructure.
- Ensuring compatibility with data pipelines.

- Establishing monitoring systems to track performance and make adjustments as necessary.
- Performing necessary training and preparing the organization to manage a post-deployment environment including marketing, legal, user training and support, and all the typical considerations when delivering a new product to market.

**Develop Analytic Code and Report:**

As part of the algorithm development phase, the data scientist is working to build an analytical system, building features, exploring the data, and ultimately, working to finalize the deployable algorithmic process.

To summarize the work, there should be an output reporting the results from all the previous steps. This report should contain information about the sample and the data, parameters, classes, methods, references and other information to help propose that the research was scientific and viable, and the algorithm works as presented. Additionally, it should be easy to follow for others so that subject matter experts can advise along the way and present a clear and comprehensive report that includes all the established metrics from the previous phases. The report should also contain easy-to-follow results explaining the responsibility metrics with a summary of the risks outlined and how the proposed model affects end users, the organization, and the costs involved. It should demonstrate insights into important distributions and samples found within the data in a question/answer format. For example: Would people in different age groups be affected differently (presented with supported visualization)? How costly are the false positives? The analysis should conclude with a clear recommendation on whether or not to proceed and outline next steps.

**Stakeholder Review Go/No Go**

After completing the above, everyone should come together in a formal meeting with formal meeting documentation to review the report, analyze the results, and come to a consensus on whether or not the work is "good enough" to proceed to deployment. If not, then create updated metrics to start another iteration of the assessment process.

Additionally, it is important to include organizational leadership with the authority to make the executive decisions in this regard, the person(s) with ownership of balancing the risks of the algorithm against the needs of the business.

The Software Alliance (2021) specifies that there should be Executive Oversight. AI Developers and AI Deployers should maintain a governance framework that is backed by sufficient executive oversight. In addition to developing and approving the substance of the governance framework's policies, senior management should play an active role in overseeing the company's AI product development lifecycle. For high-risk systems that may negatively impact people in consequential ways, company leadership should be accountable for making "go/no-go" decisions (The Software Alliance, 2021).

### Integration Planning

Now that the inferential process has been completed and all stakeholders are aligned on the algorithm, the data scientist can work with the software engineers to lock down the deployment constraints. In some cases, this can be done before the "go/no-go" decision if inputs and outputs will not change significantly. It is important to have these completed before proceeding but it is also common to have these assigned earlier in the process. It is good practice to perform integration planning as early as possible.  However,  performing this too early can create conflicts when the inputs and outputs change during the research phases.  It's best to perform this planning phase when there is good certainty that the inputs and outputs are well defined and not likely to change.

> **Topics:**
> 
> Input//Output Contract: The data engineering process will need to know what data to create to feed into the algorithm and how often, while also asking for the algorithm outputs in return. This contract can be version controlled to say: the system will always give you these data in this format, and your inference engine will always give certain data and in a certain format in return. This makes updating the algorithm in the future easier, which will be discussed later.
> 
> Update Process: This should be a plan for which version of the production will integrate the first model. It is typical that an algorithm's code may change more often than the production version and should have a process to deploy updates to the algorithm quickly without waiting for a full product release.

### Develop AI Module

Now it is time to integrate the model(s) as a new feature within the home application's development environment that is now aligned with a planned production release. At this point, the work follows typical SDLC processes where the algorithms are engineered into the overall product architecture and involves performing all the other typical processes involved with a new feature release such as marketing, educating stakeholders, etc.

### Integration

Once the code is ready, developers move it into production following the standard SDLC. It is essential to recognize that ML models and the broader application often have different deployment schedules, requiring a well-orchestrated Continuous Integration and Continuous Delivery (CI/CD) process.

For example, ML models may need to be retrained and updated more frequently than full product releases. This raises important considerations, such as how to introduce new parameters or model updates without disrupting the existing environment.

Ensuring a smooth deployment process requires clear coordination between the data science and software development teams. These discussions should focus on versioning strategies, model refresh cycles, and seamless integration workflows to maintain system stability while enabling continuous improvements.

## SAFE-AI: Part IV- Monitoring and Evaluation

After deployment, system monitoring and evaluation at regular intervals can help ensure the prevention of potential harm creation or perpetuation. While initial model results may meet responsibility metrics during development, changes in data patterns or underlying system characteristics could cause deviations in production performance.

According to the Data Science Alliance (2024), four important types of concerns exist when conducting monitoring:

- **Concept drift:** Also known as model drift, this occurs when the patterns or relationships estimated by the model shift from their state at deployment. These shifts can manifest instantaneously, gradually, cyclically, or temporarily, potentially compromising the model's effectiveness.
- **Data drift:** This phenomenon emerges when the distribution of production data significantly differs from the training data. While the negative impact of data drift on model performance typically develops slowly, its effects compound over time, leading to increasingly unreliable results.
- **Data integrity:** Production environments may introduce unexpected variables, incompatible data types, or inconsistent measurements that differ from the model's training environment. These integrity issues can severely impact the model's accuracy and reliability, requiring immediate attention and correction.
- **Bias drift:** Even a model developed with careful consideration for fairness and bias mitigation can become unfair in production due to bias drift. As societies evolve, existing biases may diminish while new ones emerge, requiring continuous monitoring and adjustment to maintain ethical performance.

The team will benefit from establishing robust evaluation frameworks to effectively monitor and address these potential drifts. The foundation of this framework relies on the fairness and transparency metrics established during model development, which serve as critical benchmarks for ongoing evaluation. By systematically tracking these metrics and implementing structured monitoring processes, it is possible to detect and respond to performance degradation before it significantly impacts stakeholders.

Below are the recommended steps to implement an effective monitoring structure:

1. **Automate Monitoring and Alerting:** Deploy automated tools that continuously track fairness and transparency metrics. Integrate these tools with alert systems to flag deviations as soon as they occur. This will enable the team to respond rapidly to potential issues, minimize system downtime, and prevent cascading failures that could affect downstream processes.
2. **Scheduled Reviews:** Implement periodic review sessions (e.g., monthly or quarterly) involving interdisciplinary teams - data scientists, clinicians, and ethics experts - to analyze monitoring reports and determine if intervention (such as model retraining or recalibration) is needed.
3. **Document Changes:** Maintain a log of all monitoring outcomes, detected drifts, interventions, and the rationale behind any model updates. This documentation is crucial for auditability and regulatory compliance.
4. **User Feedback Integration:** Establish channels for end-user (clinician and patient) feedback regarding both model performance and explainability. Use structured surveys and focus groups to gather qualitative data that can inform further improvements.
5. **Iterative Improvement:** Based on audit findings and feedback, implement retraining or fine-tuning cycles that re-incorporate recent data. Validate updated models against the established responsibility metrics before redeployment.

Effective system monitoring and evaluation represent crucial components of responsible AI deployment. Potential issues can be identified and addressed early on with the right tools in place. This proactive approach, combined with structured feedback mechanisms and continuous improvement processes, helps ensure that AI systems remain both technically sound and ethically aligned with their intended purposes.

# Discussion

## Balancing Ethical Oversight with Business Objectives

A key differentiator of the SAFE-AI ethical development framework presented in this paper is its deliberate and practical balance between ethical rigor and business priorities. Many existing frameworks, such as ECCOLA and others emerging from large academic institutions or regulatory bodies, are comprehensive but often prescriptive and cumbersome. Though they are also attempting to be agile and balance speed, they are still perceived as processes that weigh down innovation without providing a clearly articulated narrative that explains their relevance in practical business terms. Specifically, these frameworks offer valuable guidance on how to perform ethical assessments but often fail to articulate why these processes matter to the business - particularly in terms of tangible risks, operational costs, and long-term liabilities. As a result, ethical review is frequently treated as an external compliance burden rather than as an integral part of product strategy.

SAFE-AI seeks to address this gap by explicitly aligning ethical evaluation with business priorities, development speed, and stakeholder objectives. It frames ethical assessment not as an overhead cost, but as a strategic risk mitigation process that helps ensure model integrity, maintain user trust, and prevent potential regulatory, reputational, or financial harm. SAFE-AI is purposefully designed to complement the natural rhythms of modern software and AI product development by integrating ethical review seamlessly into existing Agile and Scrum workflows.

One of the primary distinctions of SAFE-AI is its emphasis on efficiency without compromising ethical standards. Rather than requiring exhaustive checklists or over-engineered governance structures, it establishes the minimum necessary assessment standards needed to preserve ethical responsibility. This allows small and medium-sized organizations to apply ethical oversight that is proportional to their product scope and business context, without overextending limited resources or slowing down development cycles.

SAFE-AI further distinguishes itself by introducing the scenario-based analogies SPAMM as an approach to communicating the ethical risks associated with AI models.Traditional interpretability tools, such as SHAP and LIME, focus on explaining the internal mechanics of models in highly technical terms. While valuable, these tools often fail to communicate the practical implications of model behavior to non-technical stakeholders, such as business leaders, clinicians, or regulatory bodies. This method addresses this gap by mapping model outputs to real-world scenarios, probabilities, and cost-based analogies that can be easily understood by multidisciplinary teams. This method facilitates more transparent discussions about risk, ethical trade-offs, and decision-making costs, making it easier for organizations to weigh these factors against business priorities.

Another key feature of the SAFE-AI framework is its emphasis on iterative ethical oversight. It recognizes that AI models are not static artifacts but evolve over time through retraining, fine-tuning, and real-world feedback. Accordingly, SAFE-AI mandates that previously defined acceptance criteria and responsibility metrics be revisited - not recreated - after any substantive changes to the model. This ensures that updates do not inadvertently introduce new ethical risks or exacerbate existing biases, while avoiding unnecessary rework and preserving business agility.

Finally, the SAFE-AI framework operationalizes ethical responsibility as a shared duty across cross-functional teams, rather than the exclusive domain of specialized ethics committees. It encourages close collaboration between technical staff, product managers, clinicians, business leaders, and other stakeholders to ensure that ethical considerations are grounded in both technical feasibility and business reality.

In summary, the SAFE-AI framework is distinct from existing methodologies in that it is practical, business-aware, and scalable. It provides a clear, actionable structure for integrating ethical best practices into AI development without compromising speed or competitiveness. By framing ethical assessment as both a product integrity safeguard and a strategic business asset, the framework empowers organizations to build responsible AI systems that are not only compliant and fair but also commercially viable and sustainable.

## The Critical Role of Risk Interpretability and Metric Development

A core principle of ethical AI development in healthcare is ensuring that the inherent risks of automated decision-making are both interpretable and measurable. Modeling risk interpretability is not a secondary concern but a foundational necessity for responsible AI deployment. Without clear and agreed-upon mechanisms for understanding how an algorithm's predictions translate into real-world outcomes, healthcare providers and affected individuals are left to navigate an opaque and potentially harmful system.

The introduction of the SAFE-AI framework addresses this challenge directly by developing probability analogy methods to bridge the gap between complex model behavior and the practical, human-driven ethical decision-making processes within healthcare settings. Unlike traditional explainability tools, which often focus exclusively on mathematical metrics or technical explanations, while emphasizing narrative-driven, probabilistic scenarios that map model behavior to understandable analogies. This approach acknowledges the reality that non-technical stakeholders, including clinicians, patients, and regulatory bodies, must ultimately be able to evaluate and engage with the outputs of AI systems in order to trust them.

The significance of this method is especially relevant during the Assessment phase of SAFE-AI. Setting responsibility metrics is not merely a technical exercise but an ethical obligation. Without these metrics, it is impossible to contextualize and evaluate how well the algorithm is performing against real-world risks. The development of responsibility metrics, including fairness and transparency metrics, must occur in tandem with risk interpretation strategies to ensure that metrics are not abstract or arbitrary but rather aligned with the actual impact on the user population.

Additionally, one of the most overlooked sources of ethical risk in AI systems is introduced not during the initial development but during subsequent algorithmic updates, such as model retraining or hyperparameter tuning. Even seemingly minor updates can alter the balance of an algorithm in subtle ways that inadvertently introduce ethical imbalances or exacerbate existing disparities. As described in the section titled "Post-Deployment Modifications," SAFE-AI mandates that any update to an existing inference engine - whether through additional training data, tuning, or rebalancing - should be treated as a new project cycle. Although previously established assessment metrics can and should be reused, they must be explicitly re-evaluated to verify that any modifications have not unintentionally undermined fairness, transparency, or safety.

This approach balances the need for iterative improvement with the ethical imperative to avoid harm. Too often, pressure to deploy quickly or incrementally improve system performance results in bypassing critical ethical assessments. SAFE-AI stresses that, even in fast-paced environments, evaluating updated models against established metrics is essential. The cost of skipping this step is not merely technical debt but ethical liability, with real consequences for patient safety and equity.

Finally, it is important to emphasize that the development of these metrics and risk interpretability mechanisms cannot be left solely to technical teams. The most effective way to safeguard against ethical harm is through multidisciplinary collaboration. Risk interpretation is inherently contextual; it requires input from healthcare professionals, regulatory experts, ethicists, and product stakeholders who understand the social and operational realities in which the AI system will function. This collaborative approach ensures that metrics are not only technically valid but ethically relevant and meaningful to the people most impacted by the model's behavior.

By prioritizing transparency, risk interpretability, and ongoing evaluation using the Responsibility Metrics approaches, the framework supports the development of AI systems that are not only effective but also fair, accountable, and ethically sound.

**Validating the Framework in Real-World Healthcare AI Settings**

To assess SAFE-AI real-world relevance and impact, future work should focus on comprehensive field-testing across diverse healthcare AI applications. The empirical analysis would involve implementing the SAFE-AI framework with small and medium-sized enterprises developing various medical AI systems, including diagnostic support tools, patient outcome assessment, and wound trajectory prediction. Using a pre-post intervention design, these studies would evaluate the framework's impact on development practices, fairness outcomes, and organizational dynamics in real-world healthcare AI deployments.

Within healthcare settings, the field studies would specifically examine how the SAFE-AI framework performs across various clinical applications, where ethical considerations regarding patient care and outcomes are particularly critical. The research would investigate how effectively SAFE-AI addresses issues such as training on biased datasets, transparency requirements, and preventing provider over-reliance on AI systems across different clinical contexts. By studying implementation across diverse medical specialties, it is possible to assess SAFE-AI's adaptability to varying clinical workflows, data types, and ethical priorities.

The proposed implementation process would include baseline assessment to capture pre-framework metrics, structured implementation of the responsibility protocols, and post-implementation evaluation at multiple intervals. This systematic approach would allow for the identification of potential adoption challenges, including resource constraints, workflow disruptions, and organizational resistance. Field testing would evaluate various implementation strategies, such as gradual integration, iterative refinement, and leadership engagement models, to determine which approaches most effectively support framework adoption while maintaining rigorous ethical standards.

Building on these future validation efforts, the framework presented here offers a practical approach to embedding ethical considerations throughout healthcare AI development while maintaining efficient workflows. By transforming abstract principles into concrete processes with measurable outcomes, this framework helps bridge the gap between technical implementation and ethical governance. As healthcare AI systems become increasingly integrated into clinical decision-making, such structured approaches to responsible development will be essential for building trustworthy systems that genuinely advance patient care without introducing new forms of bias or harm.

# Conclusion

The SAFE-AI ethical development framework proposed in this paper offers a practical, business-aligned approach for integrating responsible AI practices into the everyday workflows of product and data science teams. By mirroring established Agile and Scrum methodologies, SAFE-AI avoids treating ethics as a parallel or external process, instead embedding ethical evaluation directly into the structure of software and AI development. This alignment ensures that ethical considerations are addressed without compromising delivery timelines or development agility - a critical factor for small and medium-sized enterprises operating under resource constraints.

Central to SAFE-AI is the principle that AI development introduces uncertainty and risk that cannot be adequately managed through traditional software engineering approaches alone. Accordingly, SAFE-AI emphasizes the early and ongoing use of measurable ethical metrics - Acceptance Criteria, Fairness Metrics, and Transparency Metrics - to evaluate model behavior. These metrics serve as practical tools for managing the probabilistic nature of AI systems, enabling teams to assess whether their models meet not just performance expectations, but also ethical and social obligations.

Importantly, the SAFE-AI framework extends beyond initial deployment by incorporating a Monitoring and Evaluation phase that supports continuous ethical oversight. It recognizes that AI systems are dynamic and must be re-evaluated whenever retraining, tuning, or environmental shifts occur. This ensures that ethical integrity is not treated as a static checkpoint, but rather as an evolving requirement that must be actively maintained throughout the product lifecycle.

A key innovation introduced by SAFE-AI is the scenario-based probability analogy mapping process as an interpretability technique designed to enhance trust among non-technical stakeholders. Translating probabilistic model behavior into familiar, narrative-based analogies with associated risks and costs helps explain how and why AI systems reach certain predictions - especially in high-stakes healthcare settings where transparency is critical. Unlike traditional explainability tools that focus on model internals, SAFE-AI centers on real-world interpretability, offering an accessible way to communicate model behavior, associated risks, and uncertainty in decision-making. As such, it not only supports transparency but serves as a bridge between ethical modeling and user-centered design.

Taken together, the contributions of SAFE-AI establish a scalable pathway for responsible AI adoption in healthcare and beyond - where modeling excellence, software best practices, and ethical integrity are not at odds, but rather deeply interwoven into a cohesive and actionable development process.

# Acknowledgements

This study was partially supported by the NIA's Artificial Intelligence and Technology Collaboratories for Aging Research Grant P30AG073104 (W.W.P.); the U.S. Army xTech AI Grand Challenge (W.W.P.); the Huntsman Mental Health Foundation (W.W.P.); as well as Mountain Biometrics, Inc. (W.W.P.); Nemsee, LLC (I.N.); and the Data Science Alliance (A.M., L.J., R.L., P.L.).

The views and conclusions contained herein are those of the authors and should not be interpreted as necessarily representing the official policies, either expressed or implied, of the U.S. Army or the U.S. Government.

# Conflict of Interest

W.W.P. is a shareholder in Mountain Biometrics Inc., and I.N. is a shareholder in Nemsee, LLC. All other authors declare no competing interests.

# References

Aamodt, A., & Plaza, E. (1994). Case-Based Reasoning: Foundational Issues, Methodological Variations, and System Approaches. *AI Communications*, 7(1), 39–59. https://doi.org/10.3233/AIC-1994-7104

Abujaber, A. A., & Nashwan, A. J. (2024). Ethical framework for artificial intelligence in healthcare research: A path to integrity. *World Journal of Methodology*, 14(3), 94071. https://doi.org/10.5662/wjm.v14.i3.94071

Ali, S. J., Christin, A., Smart, A., & Katila, R. (2023). Walking the Walk of AI Ethics: Organizational Challenges and the Individualization of Risk among Ethics Entrepreneurs. *Proceedings of the 2023 ACM Conference on Fairness, Accountability, and Transparency*, 217–226. https://doi.org/10.1145/3593013.3593990

Atlassian. (n.d.). *Acceptance Criteria Explained*. Atlassian. Retrieved January 23, 2025, from https://www.atlassian.com/work-management/project-management/acceptance-criteria

Babic, B., Gerke, S., Evgeniou, T., & Cohen, I. G. (2021). Beware explanations from AI in health care. *Science*, *373*(6552), 284–286. https://doi.org/10.1126/science.abg1834

Baquero, J. A., Burkhardt, R., Govindarajan, A., & Wallace, T. (2020, August 13). *Derisking AI: Risk management in AI development | McKinsey*. https://www.mckinsey.com/capabilities/quantumblack/our-insights/derisking-ai-by-design-how-to-build-risk-management-into-ai-development?utm_source=chatgpt.com

Berman, G., Goyal, N., & Madaio, M. (2024). A Scoping Study of Evaluation Practices for Responsible AI Tools: Steps Towards Effectiveness Evaluations. *Proceedings of the CHI Conference on Human Factors in Computing Systems*, 1–24. https://doi.org/10.1145/3613904.3642398

Bevilacqua, M., Berente, N., Domin, H., Goehring, B., & Rossi, F. (2023). *The Return on Investment in AI Ethics: A Holistic Framework* (arXiv:2309.13057). arXiv. https://doi.org/10.48550/arXiv.2309.13057

Cai, C. J., Jongejan, J., & Holbrook, J. (2019). The effects of example-based explanations in a machine learning interface. *Proceedings of the 24th International Conference on Intelligent User Interfaces*, 258–262. https://doi.org/10.1145/3301275.3302289

Centers for Medicare & Medicaid Services. (2024, May 6). *Nondiscrimination in Health Programs and Activities*. https://www.federalregister.gov/documents/2024/05/06/2024-08711/nondiscrimination-in-health-programs-and-activities

Coalition for Health AI. (2023, April 4). *Blueprint for Trustworthy AI Implementation Guidance and Assurance for Healthcare Version 1.0*. https://chai.org/wp-content/uploads/2024/05/blueprint-for-trustworthy-ai_V1.0-2.pdf

Costa, M. (2013). Social Return on Investment. In *Encyclopedia of Corporate Social Responsibility* (pp. 2238–2248). https://doi.org/10.1007/978-3-642-28036-8_734

Crockett, K., Colyer, E., Gerber, L., & Latham, A. (2023). Building Trustworthy AI Solutions: A Case for Practical Solutions for Small Businesses. *IEEE Transactions on Artificial Intelligence*, *4*(4), 778–791. IEEE Transactions on Artificial Intelligence. https://doi.org/10.1109/TAI.2021.3137091

Data Science Alliance. (2024, February). *The Framework for Responsible Data Science Practices*. https://drive.google.com/file/d/17cdW49APhyv7qEy0EyQV6vv7qkw_EcIQ/view?usp=embed_facebook

Domin, H., Rossi, F., Goehring, B., Ganapini, M., & Berente, N. (n.d.). *2024-07-on-the-roi-of-ai-ethics-and-governance-investments-from-loss-aversion-to-value-generation*.

Fairlearn. (n.d.). *Assessment—Fairlearn 0.12.0 documentation*. Retrieved January 23, 2025, from https://fairlearn.org/v0.12/user_guide/assessment/index.html

Fjeld, J., Achten, N., Hilligoss, H., Nagy, A., & Srikumar, M. (2020). *Principled Artificial Intelligence: Mapping Consensus in Ethical and Rights-Based Approaches to Principles for AI* (SSRN Scholarly Paper 3518482). Social Science Research Network. https://doi.org/10.2139/ssrn.3518482

Hajianhosseinabadi, H., & Lindau, N. (2024). *Startup interaction with AI - A multiple case study on startups' ethical navigation when using Generative AI tools*. https://gupea.ub.gu.se/handle/2077/82378

Health, C. for D. and R. (2023). Total Product Life Cycle for Medical Devices. *FDA*. https://www.fda.gov/about-fda/cdrh-transparency/total-product-life-cycle-medical-devices

IONOS Editorial Team. (2025, February 3). *What is case-based reasoning?* IONOS Digital Guide. https://www.ionos.com/digitalguide/websites/web-development/case-based-reasoning/

Kim, S. (Josh). (2022, January 19). *Explainable AI (XAI) Methods Part 3—Accumulated Local Effects (ALE)*. Towards Data Science. https://towardsdatascience.com/explainable-ai-xai-methods-part-3-accumulated-local-effects-ale-cf6ba3387fde/

Kostick-Quenet, K. M., Cohen, I. G., Gerke, S., Lo, B., Antaki, J., Movahedi, F., Njah, H., Schoen, L., Estep, J. E., & Blumenthal-Barby, J. S. (2022). Mitigating Racial Bias in Machine Learning. *The Journal of Law, Medicine & Ethics: A Journal of the American Society of Law, Medicine & Ethics*, *50*(1), 92–100. https://doi.org/10.1017/jme.2022.13

Mayo. (2022, March 23). The AI/ML Wars: "Explain" or test black box models? *Error Statistics Philosophy*. https://errorstatistics.com/2022/03/23/the-ai-ml-wars-explain-or-test-black-box-models/

Mello, M. M., & Guha, N. (2024). Understanding Liability Risk from Healthcare AI. *Policy Brief, Stanford University Human-Centered Artificial Intelligence*. https://hai-production.s3.amazonaws.com/files/2024-02/Liability-Risk-Healthcare-AI.pdf

Millar, R., & Hall, K. (2013). Social Return on Investment (SROI) and Performance Measurement: The opportunities and barriers for social enterprises in health and social care. *Public Management Review*, *15*(6), 923–941. https://doi.org/10.1080/14719037.2012.698857

Molnar, C. (n.d.). *Chapter 6 Model-Agnostic Methods | Interpretable Machine Learning*. Retrieved February 5, 2025, from https://christophm.github.io/interpretable-ml-book/agnostic.html

Mutegeki, H., Nahabwe, A., Nakatumba-Nabende, J., & Marvin, G. (2023). Interpretable Machine Learning-Based Triage For Decision Support in Emergency Care. *2023 7th International Conference on Trends in Electronics and Informatics (ICOEI)*, 983–990. https://doi.org/10.1109/ICOEI56765.2023.10125918

Ning, Y., Liu, X., Collins, G. S., Moons, K. G. M., McCradden, M., Ting, D. S. W., Ong, J. C. L., Goldstein, B. A., Wagner, S. K., Keane, P. A., Topol, E. J., & Liu, N. (2024). An ethics assessment tool for artificial intelligence implementation in healthcare: CARE-AI. *Nature Medicine*, *30*(11), 3038–3039. https://doi.org/10.1038/s41591-024-03310-1

NIST AIRC. (n.d.). *NIST AIRC - Playbook*. Retrieved January 23, 2025, from https://airc.nist.gov/AI_RMF_Knowledge_Base/Playbook

Pfeiffer, R. S. (1992). Teaching Ethical Decision-Making: The Case Study Method and the RESOLVEDD Strategy. *Teaching Philosophy*, *15*(2), 175–184. https://doi.org/10.5840/teachphil199215221

Pfeiffer, R. S., & Forsberg, R. P. (2014). *Ethics on the Job: Cases and Strategies, Fourth Edition*. Wadsworth, Cengage Learning.

Ratwani, R. M., Sutton, K., & Galarraga, J. E. (2024). Addressing AI Algorithmic Bias in Health Care. *JAMA*, *332*(13), 1051–1052. https://doi.org/10.1001/jama.2024.13486

Reddy, G. P., Pavan Kumar, Y. V., & Prakash, K. P. (2024). Hallucinations in Large Language Models (LLMs). *2024 IEEE Open Conference of Electrical, Electronic and*

*Information Sciences (eStream)*, 1–6. https://doi.org/10.1109/eStream61684.2024.10542617

Rep. Arrington, J. C. [R-T.-19. (2025, June 29). *Text - H.R.1 - 119th Congress (2025-2026): One Big Beautiful Bill Act* (2025-05-20) [Legislation]. https://www.congress.gov/bill/119th-congress/house-bill/1/text

Ribeiro, M. T., Singh, S., & Guestrin, C. (2016). *Model-Agnostic Interpretability of Machine Learning* (arXiv:1606.05386). arXiv. https://doi.org/10.48550/arXiv.1606.05386

Saenz, A. D., Centi, A., Ting, D., You, J. G., Landman, A., & Mishuris, R. G. (2024). Establishing responsible use of AI guidelines: A comprehensive case study for healthcare institutions. *Npj Digital Medicine*, *7*(1), 1–6. https://doi.org/10.1038/s41746-024-01300-8

Scrum Alliance. (n.d.). *Acceptance Criteria: Everything You Need to Know Plus Examples*. Retrieved January 23, 2025, from https://resources.scrumalliance.org/Article/need-know-acceptance-criteria

The Software Alliance, B. (2021). *Confronting Bias: BSA's Framework to Build Trust in AI*. https://www.bsa.org/files/reports/2021bsaaibias.pdf

US Department of Health and Human Services. (n.d.). *Informed Consent FAQs*. Retrieved February 5, 2025, from https://www.hhs.gov/ohrp/regulations-and-policy/guidance/faq/informed-consent/index.html

US Food and Drug Administration. (2022, September 28). *Clinical Decision Support Software Guidance for Industry and Food and Drug Administration Staff*. https://www.fda.gov/media/109618/download

US Food and Drug Administration. (2023, October 20). *Good Machine Learning Practice for Medical Device Development: Guiding Principles*. FDA. https://www.fda.gov/medical-devices/software-medical-device-samd/good-machine-learning-practice-medical-device-development-guiding-principles

US Food and Drug Administration. (2025a, January 6). *Artificial Intelligence and Machine Learning in Software as a Medical Device*. FDA. https://www.fda.gov/medical-devices/software-medical-device-samd/artificial-intelligence-and-machine-learning-software-medical-device

US Food and Drug Administration. (2025b, January 8). *Clinical Decision Support Software Frequently Asked Questions (FAQs)*. FDA. https://www.fda.gov/medical-devices/software-medical-device-samd/clinical-decision-support-software-frequently-asked-questions-faqs

Vakkuri, V., Kemell, K.-K., & Abrahamsson, P. (2019). Ethically Aligned Design: An Empirical Evaluation of the RESOLVEDD-Strategy in Software and Systems Development Context. *2019 45th Euromicro Conference on Software Engineering and Advanced Applications (SEAA)*, 46–50. https://doi.org/10.1109/SEAA.2019.00015

Vakkuri, V., Kemell, K.-K., Jantunen, M., Halme, E., & Abrahamsson, P. (2021). ECCOLA — A method for implementing ethically aligned AI systems. *Journal of Systems and Software*, *182*, 111067. https://doi.org/10.1016/j.jss.2021.111067

Watson, D. (2020). Conceptual Challenges for Interpretable Machine Learning. *SSRN Electronic Journal*. https://doi.org/10.2139/ssrn.3668444

Weerts, H., Dudík, M., Edgar, R., Jalali, A., Lutz, R., & Madaio, M. (2023). *Fairlearn: Assessing and Improving Fairness of AI Systems* (arXiv:2303.16626). arXiv. https://doi.org/10.48550/arXiv.2303.16626

Winecoff, A. A., & Watkins, E. A. (2022). Artificial Concepts of Artificial Intelligence: Institutional Compliance and Resistance in AI Startups. *Proceedings of the 2022*

*AAAI/ACM Conference on AI, Ethics, and Society*, 788–799. https://doi.org/10.1145/3514094.3534138